\pdfoutput=1
\documentclass{sig-alt-gov2}
\usepackage{verbatim, pseudocode, times,amsmath,epsfig,algorithm2e}
\title{NOUS: Construction and Querying of Dynamic Knowledge Graphs}
\author{%
Sutanay Choudhury{\small $~^{\#1}$}
Khushbu Agarwal{\small $~^{\#1}$}
Sumit Purohit{\small $~^{\#1}$}
Baichuan Zhang{\small $~^{\#2}$} \\
Meg Pirrung{\small $~^{\#1}$}
Will Smith{\small $~^{\#1}$}
Mathew Thomas{\small $~^{\#1}$}
\vspace{1.6mm}
\fontsize{10}{10}\selectfont\itshape
$^{\#1}$\, Pacific Northwest National Laboratory,\\
902 Battelle Blvd, Richland, WA 99352\\
$^{\#2}$\, Purdue University,\\
Indianapolis, IN\\
}
\begin{document}
\maketitle
\begin{abstract}
The ability to construct domain specific knowledge graphs (KG) and perform question-answering or hypothesis generation is a transformative capability.  Despite their value, automated construction of knowledge graphs remains an expensive technical challenge that is beyond the reach for most enterprises and academic institutions.  We propose an end-to-end framework for developing custom knowledge graph driven analytics for arbitrary application domains.  The uniqueness of our system lies A) in its combination of curated KGs along with knowledge extracted from unstructured text, B) support for advanced trending and explanatory questions on a dynamic KG, and C) the ability to answer queries where the answer is embedded across multiple data sources.
\end{abstract}
\section{Introduction}
\label{sec:intro}

Data-driven applications critically depend on human experts who learn about a given domain through their interaction with data over time.  An expert market analyst can quickly answer questions about recently trending products or explain the reason behind a new trend.  However, manual approaches stops scaling as the data volume, throughput and diversity surges.  Consequently, organizations are building custom knowledge bases (KB) using a combination of human-in-the-loop and data-driven techniques.   These efforts range across diverse domains such as cyber-security \cite{iannacone2015developing}, medical diagnosis ~\cite{lally:ibm14} and retail \cite{deshpande:sigmod13}.

Domain specific KB construction frequently begins with knowledge extraction from unstructured data such as publications, web pages or social media.  The extracted knowledge is often aimed at improving the quality of search or recommendation algorithms.  Another relevant area is Question-Answering (QA), which involves answering natural language queries on a large-scale text corpus \cite{ferrucci:aaai10}.  Understanding domain specific vocabularies, correct classification of entities and their relationships is key to accomplishing these goals.  Therefore, building on top of curated KBs such as FreeBase, YAGO etc. and augmenting with knowledge extracted by techniques such as Open Information Extraction \cite{banko2007open} has emerged as a natural path for custom KG construction.

In short, our work is set in the following paradigm: 1) data arrives in streaming fashion and knowledge extraction happens continuously, 2) extracted knowledge is combined with curated knowledge to form a dynamic KG, and 3) a set of queries are executed on the dynamic KG.  Executing queries on the KGs provides an unique analytics perspective. Since each relationship in a KG is potentially extracted from different data sources, it allows us to connect the dots across multiple data sources.  In contrast, systems that return a passage of text from the best matching document does not provide such capabilities.

\begin{figure}[htbp]
\centering
\includegraphics[scale=0.27]{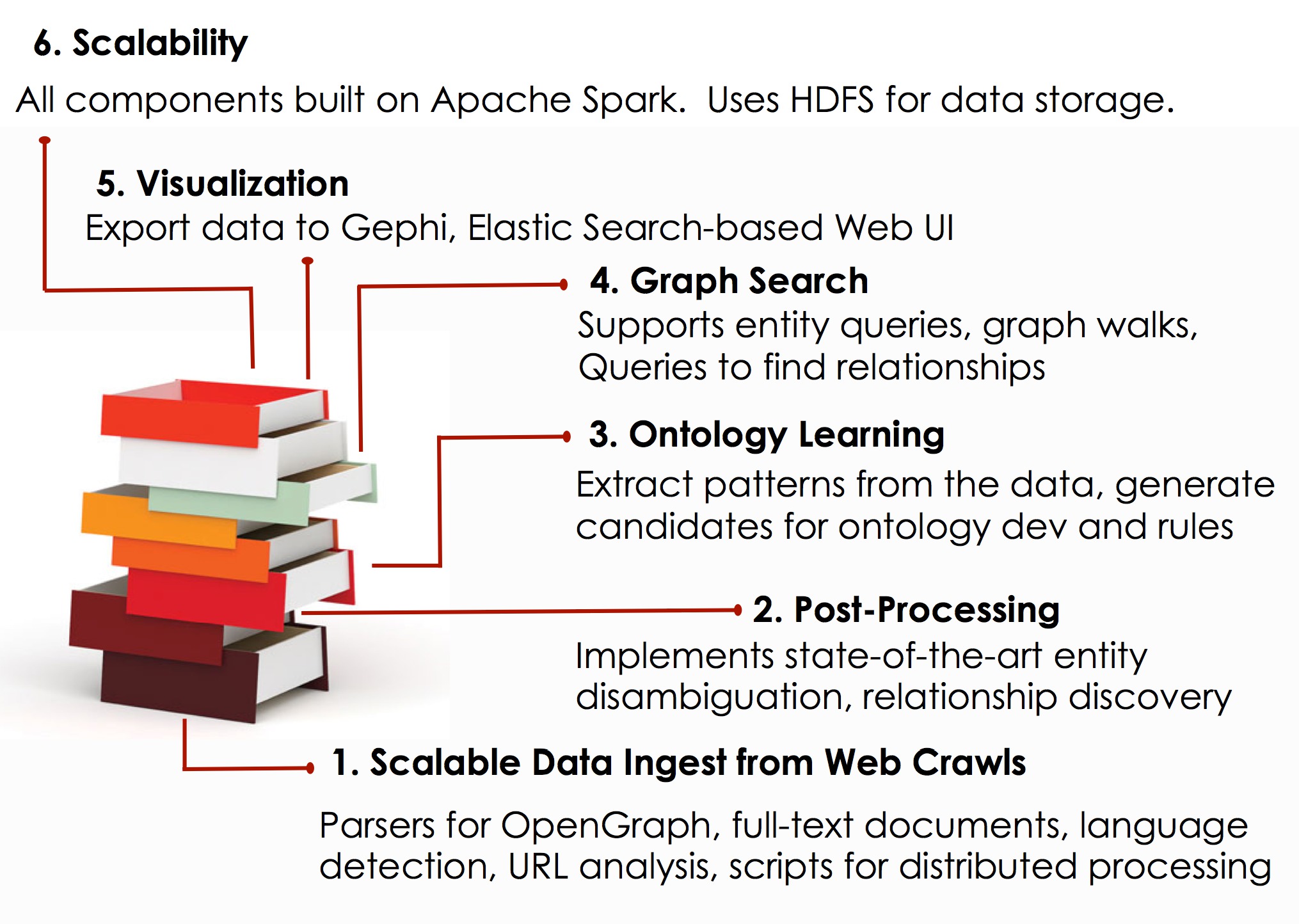}
\caption{Various components of NOUS.}
\label{fig:system}
\end{figure}

\subsection{Technical contributions}

Our proposed demonstration will showcase NOUS ~\footnote{NOUS means ``experiential knowledge" in Greek, one that is gathered over time.} (Figure \ref{fig:system}), an open source system ~\footnote{\textbf{https://github.com/streaming-graphs/NOUS}}. Following are the primary contributions of NOUS.

\begin{enumerate}
\item We develop a framework to build domain specialized knowledge graphs by fusing curated KBs with extracted knowledge.  This is in contrast to most systems who either leverage on curated KBs such as Freebase or work on extracted knowledge.  We view the Knowledge Graph construction as an incremental process and develop a family of algorithms designed for dynamic graphs.
\item We focus on two popular classes of domain-specific application needs:  1) discovering trends in streaming data and 2) answering explanatory (why-like) questions.  We implement the former via a novel algorithm for streaming graph mining (section 3.5).  We implement (2) by augmenting state of the art path-ranking algorithms with a \textsl{coherence} metric based approach (section 3.6).
\item We execute the queries on a dynamically updated Knowledge Graph.  This allows NOUS to support queries whose answers are composed from multiple data sources.
\end{enumerate}

Rest of the paper is organized as follows.  Section 1.2 illustrates a primary use case for motivation.  Section 2 provides an overview of related literature.  Section 3 describes the details of various technical components and section 4 concludes with specific features of the demonstration.

\subsection{Use Case}
\begin{figure}[h]
\centering
\includegraphics[scale=0.45]{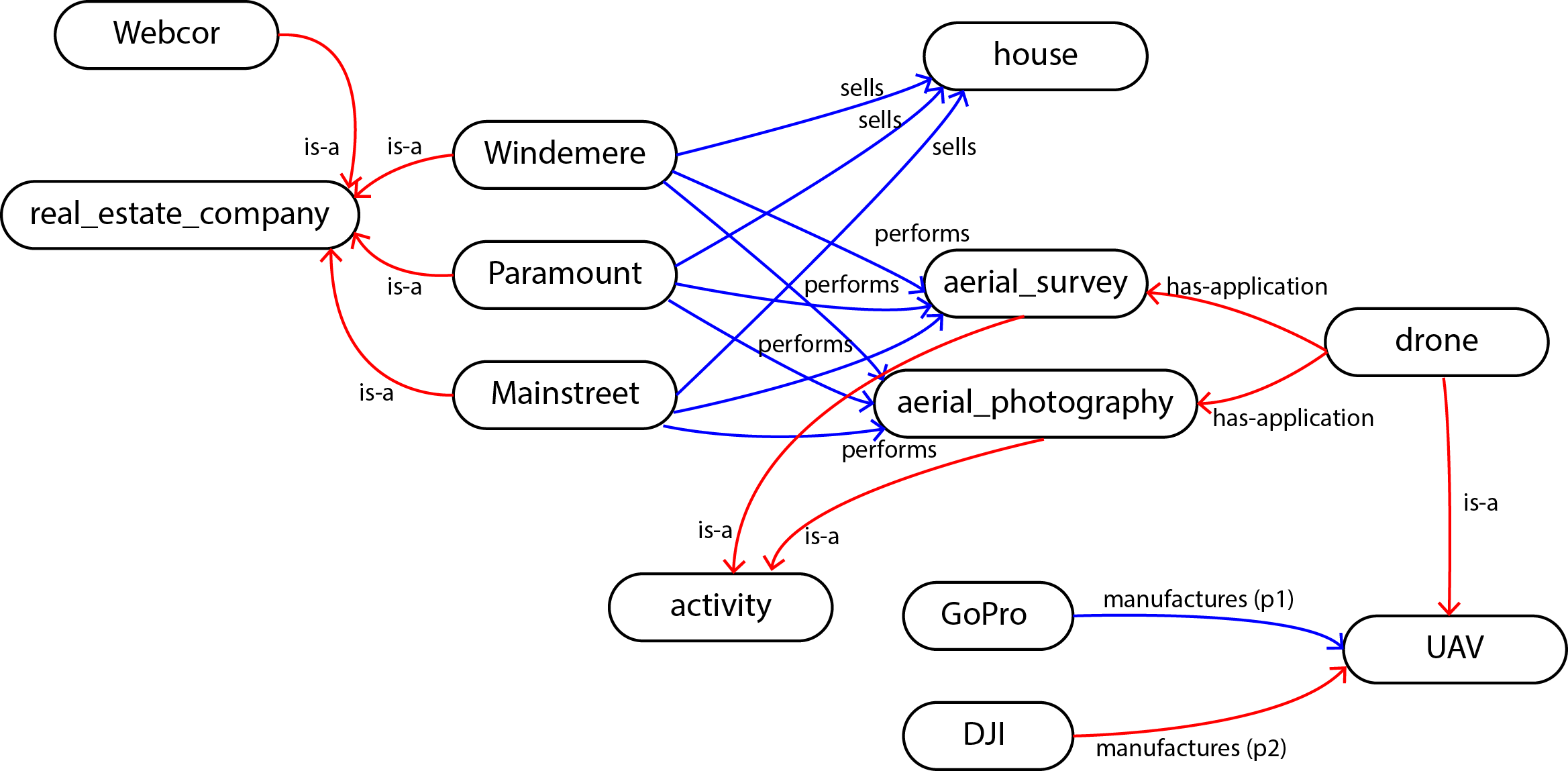}
\caption{Example of a Knowledge Graph tracking emerging technologies.}
\label{fig:droneGraph}
\end{figure}

Civilian use of Drones is a key emerging technology today.  Finance analysts, law enforcement agencies need to track emerging drone manufactures, novel applications of drones, and identify safety issues.   Our goal is to build a system that can ingest information from web crawls and articles from trusted news sources to continuously stay abreast of information around drones.  From the user perspective, an finance analyst may hypothesize about a startup being the acquisition target for a novel drone-based technology, or a security analyst will want to reason about why a non-military organization such as Windermere may employ drones in their operations (Figure \ref{fig:droneGraph}).  Figure~\ref{fig:droneGraph} shows a sample of the drone graph generated by fusing knowledge from YAGO2 and Wall street journal articles. The lines in red and blue indicate facts available from curated KB and facts learned from web data respectively. Each fact is assigned a probability value of it being true, learned using the Link Prediction module.

\section{Background and Related Work}
\label{sec:background}

Knowledge Graphs with their ability to represent complex relationships
between real world entities have become the de facto standard to store KBs.  Knowledge graph construction from web data~\cite{dong:kdd14, deshpande:sigmod13, ferrucci:aaai10, niu2012deepdive} has been studied comprehensively over the last decade. {\em Deshpande et al}~\cite{deshpande:sigmod13} provide an in depth discussion of the process and associated challenges. Openly available KBs like YAGO~\cite{suchanek:www07}, Freebase~\cite{bollacker:sigmod08} and NELL~\cite{carlson:aaai10} provide massive amount of highly confident triples. We view these general purpose KB's as complimentary, to be used in conjunction with NOUS to build a custom domain KG. Most of these KGs or frameworks are limited in their querying capabilities.  While algorithms for querying or mining dynamic graphs have been studied \cite{choudhury2015selectivity}, much of the research happened without addressing Knowledge Graph specific issues.
%Commercial systems like Google's knowledge vault~\cite{dong:kdd14}, IBM Watson~\cite{ferrucci:aaai10} are unavailable to the user community.
%DeepDive~\cite{niu2012deepdive} is system with similar objective and approach with NOUS.  Its human-in-the-loop component requires users to write rules to infer relations, whereas NOUS takes a visual analytics approach to identify meaningful relations.  Question-Answering is an important objective for NOUS, which is not a focus area for DeepDive.
\section{NOUS Overview}
\label{sec:overview}
This section describes various components of NOUS's knowledge graph construction pipeline as shown in Figure~\ref{fig:system}.  The user interface and the question-answering system is discussed in the next section.  In the text we emphasize on components where NOUS is making a research contribution and point the reader to the literature ~\cite{deshpande:sigmod13, dong:kdd14} for a thorough discussion of the knowledge graph construction process and challenges.

\subsection{Data Sources}

Algorithms in NOUS are being used for developing custom knowledge graphs for diverse domains : 1) business intelligence applications via news articles and web crawls, 2) insider threat detection using various log data sources from enterprises and 3) citation analytics from bibliography databases.  We will restrict our discussion to Wall Street Journal articles and general web crawls for the purpose of this demonstration.

\subsection{Triple Extraction from Natural Language Text}
We extract the text from every input document (such as a news article or blog post), and then process it sentence by sentence for entity and relation extraction.  For relationship extraction, we used Open Information Extraction (OpenIE) \cite{banko2007open} technique to obtain binary or n-ary relational tuples from every sentence.  We also perform named entity extraction and co-reference resolution, and used this information to implement heuristics for triple extraction.

\subsection{Mapping Raw Triples to Knowledge Graph}

Our goal is combine the extracted triples with a high-quality, openly available KG such as YAGO.  Therefore, we need to map the subjects and objects in the triples to entities present in YAGO2, or else create a new node or relation into our custom knowledge graph.

A challenge with OpenIE like techniques is that they produce too many relations. We implement a distant supervision based approach to learn a rule-based model for each predicate and map the predicates from raw triples to the target ontology.  Following the \textsl{Extreme Extraction} work by Freedman et al\cite{freedman2011extreme}, we bootstrap each predicate model with 5-10 seed examples and expand the set of training examples for each predicate in a semi-supervised fashion.  This is still an active area of refinement for NOUS.

We implement a variation of the AIDA algorithm proposed by {\em Hoffart et al}~\cite{hoffart:emnlp11} for entity disambiguation.  AIDA was chosen due to its high accuracy, scalability and ease of implementation in Spark platform.  We adapted AIDA's context based similarity score that was originally based on comparing entity's Wikipedia article to the text surrounding the entity mention.  As new entities from online articles are added to the knowledge graph, we use only the entity
neighborhood in the knowledge graph to calculate contextual similarity.

\subsection{Confidence Estimation via Link Prediction}
Triples extracted from the text data sources are extremely noisy, and simply adding noisy facts to the knowledge graph will destroy its purpose.  In addition to tracking source level trust, we implemented a Link Prediction approach \cite{zhang2016trust} to quantitatively measure confidence in a triple using the prior state of the knowledge graph.  For every predicate we build a latent feature embedding model using Bayesian Personalized Ranking (BPR) as the optimization criteria.  Given an input triple, the model produces a real-valued score between 0 and 1.

\subsection{Rule Learning via Frequent Graph Mining}
Frequent Graph Mining (FGM) is the process of finding highly frequent subgraphs that help to discover structural association between entities and functional dependencies.  A major research contribution of NOUS is the development of a distributed algorithm for streaming graph mining.  Another novelty of our implementation is its ability to simultaneously support the curated KB and the extracted knowledge, and discover patterns by combining both structures.

The algorithm accepts the stream of incoming triples as input, a window size parameter that represents the size of a sliding window over the stream and reports the set of closed frequent patterns present in the window.  As the stream characteristics change and some patterns turn from frequent to infrequent, our algorithm supports reconstruction of smaller frequent patterns from larger patterns that just turned infrequent.  Distinct from transaction setting based algorithms such as gSpan \cite{yan2002gspan}, initial benchmarking of our work against distributed graph mining systems such as Arabesque \cite{teixeira2015arabesque} suggests 3x speedup on selected datasets.

\subsection{Question Answering}
The usability of NOUS depends on its ability to answer questions of interest to domain users.  We implemented a novel path search algorithm for Knowledge Graphs. The algorithm accepts three arguments as input: a source $s$ and a target entity $t$, and a relationship constraint, which typically is a predicate from the target ontology.  Given this input, the algorithm returns a set of top-K paths to explain the relationship between $s$ and $t$.

%Traditional path search algorithms in graphs have focussed on shortest-path~\cite{Qi:vldb13} and reachability queries. We observe the limitations of such methods in finding a coherent path between entities. For e.g., shortest path approach to

Our graph search algorithm is implemented on top of the distributed property graph model available from Apache Spark's GraphX library.  The GraphX implementation allows us to store arbitrary properties with the vertices and edges.  We utilize the text datasets available for each vertex in the graph (such as bag-of-words extracted from the Wikipedia page of an entity) and assign a topic distribution to every entity by executing the Latent Dirichlet Allocation (LDA) algorithm on the ``document-term" matrix constructed from the text.  During the graph walk, we perform a look-ahead search at every hop and select nodes with least topic divergence to the target node.  Finally, we compute a ``coherence" score for every path between the source and target, and the path with least amount of divergence is chosen.

%\subsubsection{Path Query}
%The aim of a path query is to detect 'meaningful' relationships between entities of a interest,
%such as "Howz entity X related to Y".
%Graph path finding methods like shortest-distance and reachability queries have been well studied in
%the literature. However, we argue that existing methods like shortest-distance
%do not provide a contextual answer to the user.
%For e.g. a shortest distance approach to answering following path query
%
%"How Lebron James knows Draymond Green",  returns
%
%{\em Lebron James} - {\em rdf:type} - {\emMale} - {\em rdf:type} - {\em Daymond:Green}
%
%We instead implement a topic coherence based path ranking algorithm.
%We build the topic model of each yago entity from its wikipedia
%text article offline using Latent Dirichlet Allocation(LDA).
%During the graph walk we compute the topic divergence
%of source node with destination's neighbors and vice versa. This guides the walk to a topic coherent
%result. A user defined parameter defines the maximum relationship distance acceptable to the user,
%which is used to terminate the walk if no acceptable path is found.

\section{Demonstration Features: What to Expect}
\label{sec:demo}

NOUS is implemented using the Scala programming language (version 1.5) on top of Apache Spark~\footnote{http://spark.apache.org}.  Our demonstration will use a Spark cluster running inside PNNL's compute cloud.  We will use the Wall Street Journal (WSJ) corpus from 2010-2015 comprising 342,411 articles as the primary dataset and provide hands-on demonstration of the following features. The overarching goal of the demonstration will be to show how NOUS can be used to build custom knowledge graphs from web scale data and answer unique domain-specific queries.

%\begin{enumerate}
1. Develop custom relation extractors and illustrate the trade-off from various heuristics.

2. Visualize the resultant graph and summarization of quality-related statistics (such as confidence distributions, and understanding how the structure of the underlying data influence the output quality).

3. Develop custom quality control modules for a new domain.

4. Execute queries for pattern discovery and graph search using both web and command line interface.
%\end{enumerate}

%\subsection{Open Source Codebase}
%Our current codebase is available at https://github.com/streaming-graphs/NOUS.

%\begin{figure}[h]
%    \centering
%    \includegraphics[width=1.0\columnwidth]{./figs/search2}
%    \includegraphics[width=1.0\columnwidth]{./figs/search3}
%    \caption{NOUS Search Interface}
%    \label{fig:searchUI}
%\end{figure}

%As mentioned above we initialize NOUS with the YAGO2 knowledge graph due to its scale and accuracy. YAGO2 contains 447M facts about 9.8M entities and reported 95\% accuracy on its facts~\cite{hoffart:jai13} during a human evaluation. It takes NOUS nearly 12 minutes to load YAGO2 in GraphX in binary format.
%We will use the Wall Street Journal (WSJ) corpus from 2010-2015 comprising 342,411 articles to demonstrate our
%system. We extracted 18.28M triples and 364K entities. We plan to showcase the trend analysis generated using FGM module over the fused KB. For e.g:
% "technology company, manufacturer, drone" , and the pattern evolution over time. We will
%also present the confidence values generated for WSJ triples and the associated statistics.
%The User Interface(UI) will feature search capability as shown in Figure ~\ref{fig:searchUI} where they can interact
%the interface and ask queries of supported types.
%
%\begin{figure}[h]
%    \centering
%    \includegraphics[width=1.0\columnwidth]{./figs/search2}
%    \includegraphics[width=1.0\columnwidth]{./figs/search3}
%    \caption{NOUS Search Interface}
%    \label{fig:searchUI}
%\end{figure}
\section{Acknowledgements}
The research is sponsored by the Analysis in Motion Initiative at Pacific Northwest National Laboratory.  We would also like to thank the OpenKE team at Laboratory for Analytic Sciences for their continuous support and feedback.
\bibliographystyle{abbrv}
\bibliography{nous}
\newpage
\appendix

\begin{figure}[h]
\centering
\includegraphics[scale=0.2]{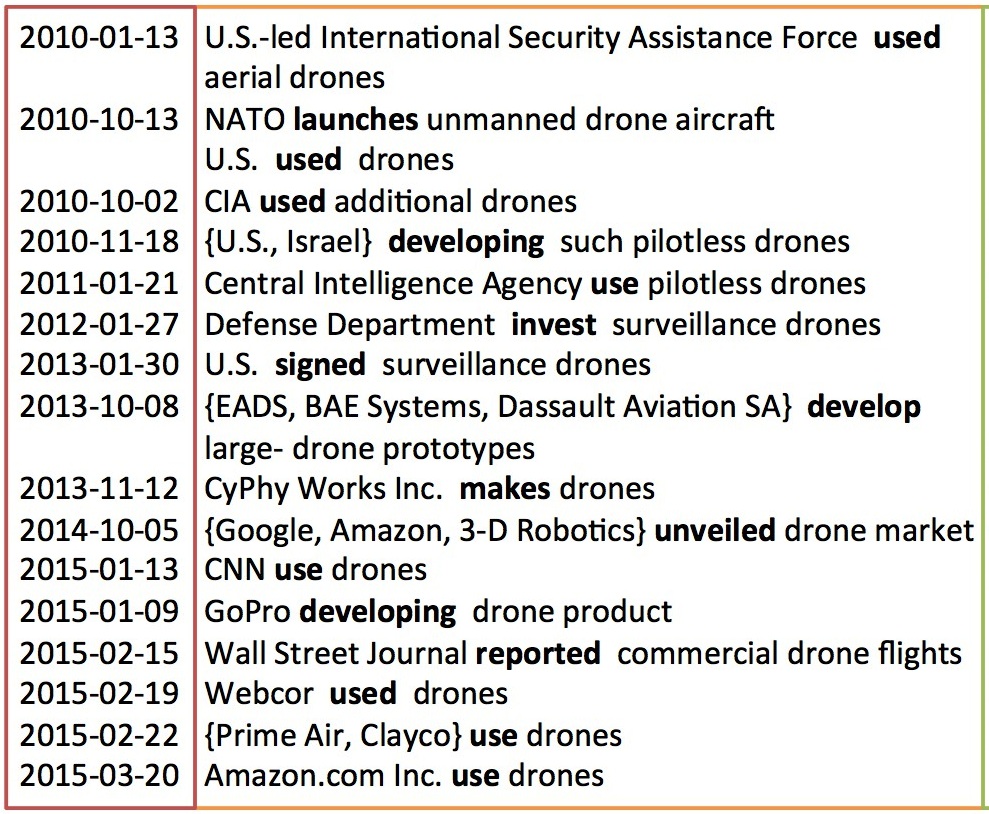}
\caption{Example triples extracted from Wall Street Journal Articles using Semantic Role Labeling.  The first column shows dates on which the triples were published.}
\label{fig:droneTriples}
\end{figure}

\begin{figure}[h]
\centering
\includegraphics[scale=0.23]{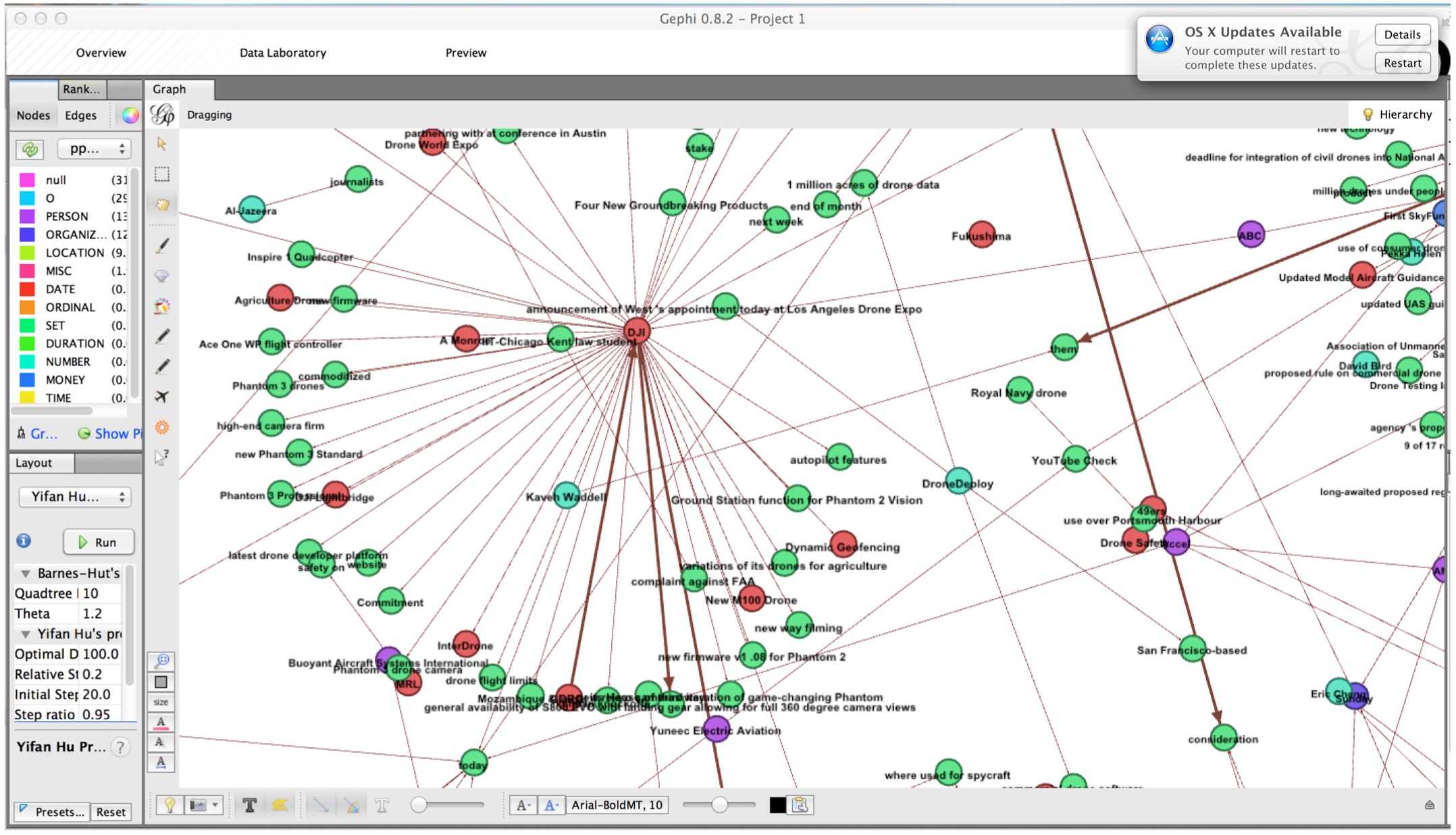}
\caption{Visualization of a subgraph from a drone-themed Knowledge Graph.  Algorithms in NOUS can be used to develop custom knowledge graphs for any arbitrary application domain.}
\label{fig:knowledge_graph}
\end{figure}

\begin{figure}[]
\centering
\includegraphics[scale=0.25]{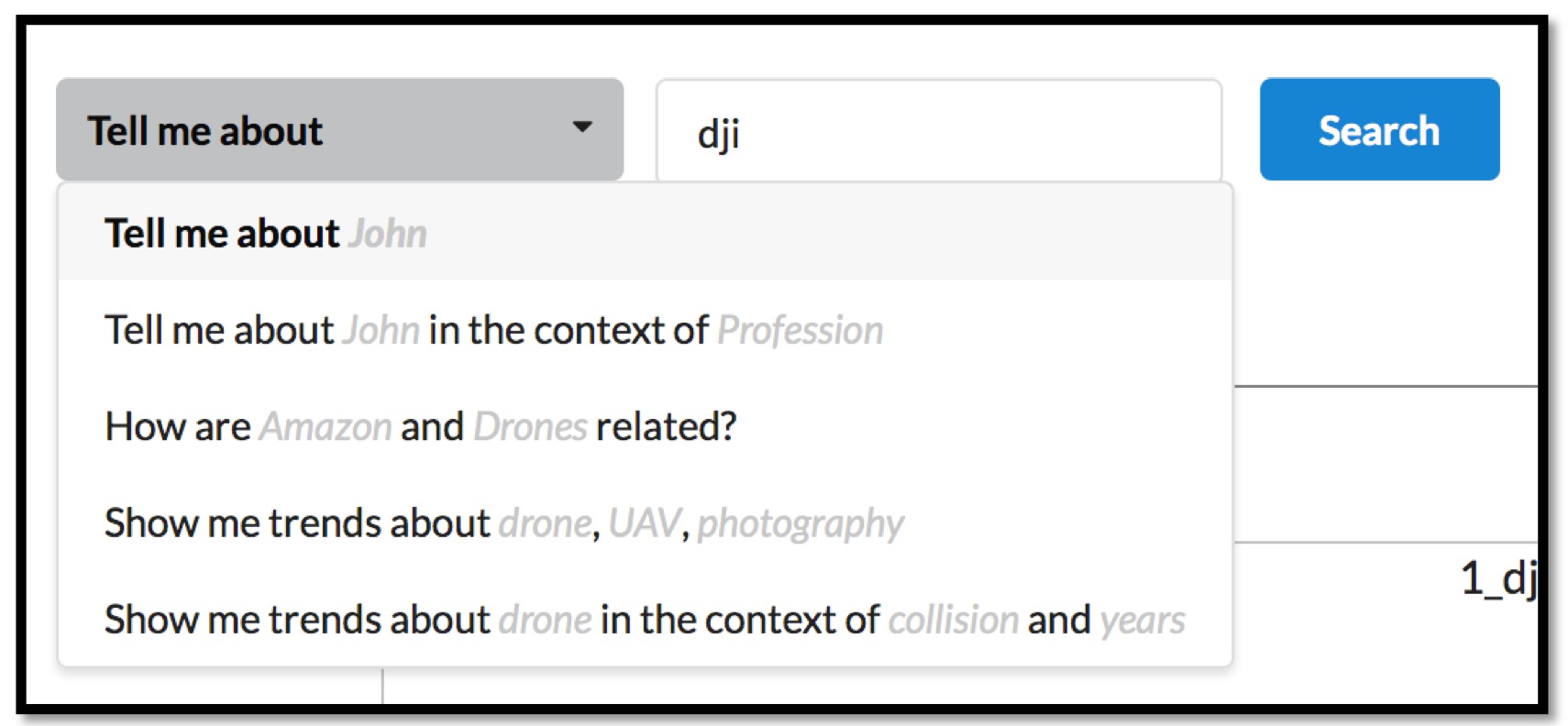}
\caption{Five classes of natural language like queries that are transparently translated to execute distributed algorithms for subgraph pattern mining, entity-based queries or complex graph queries.}
\label{fig:nlp_query_interface}
\end{figure}

\begin{figure}[h]
\centering
\includegraphics[scale=0.18]{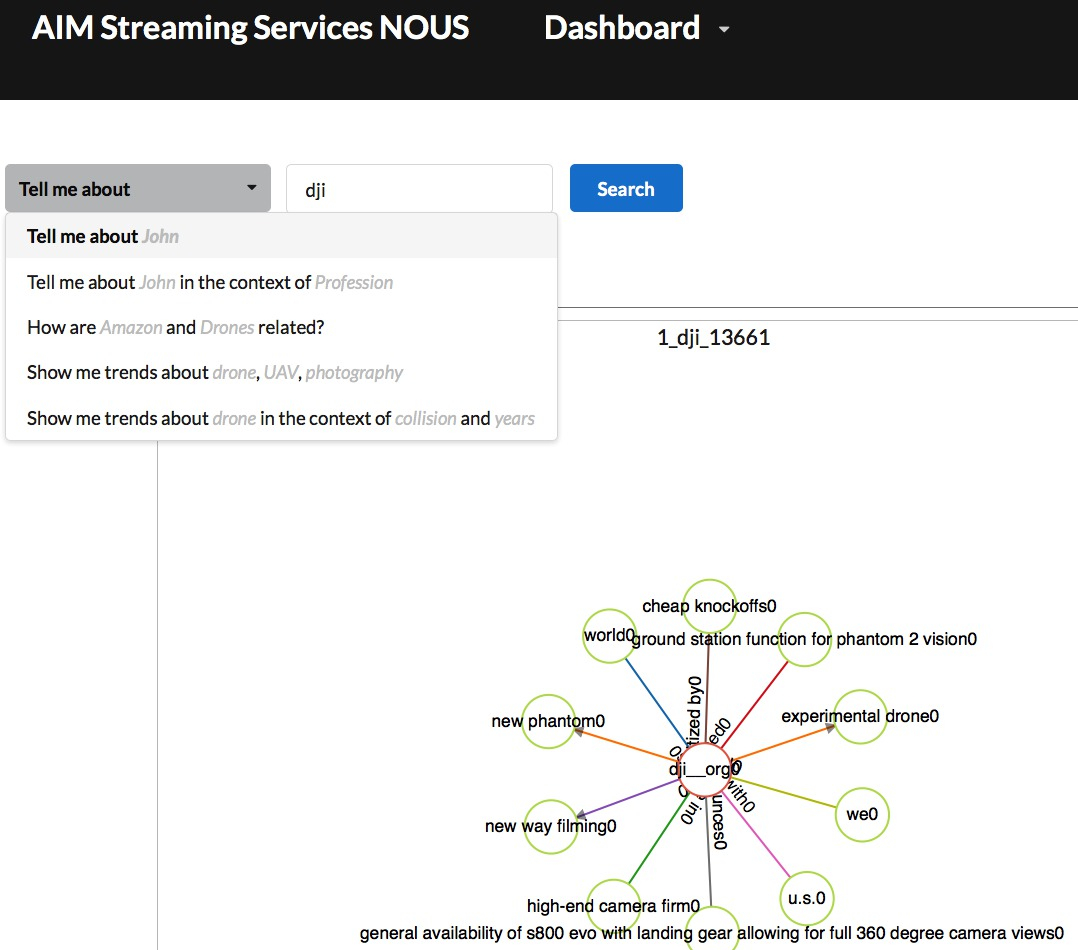}
\caption{Web based interface for Trending, Entity and Relationship-based queries. The image here shows results for an entity query ``Tell me about DJI", where DJI is world's leading company for manufacturing drones.}
\label{fig:query_interface}
\end{figure}

\begin{figure}[h]
\centering
\includegraphics[scale=0.2]{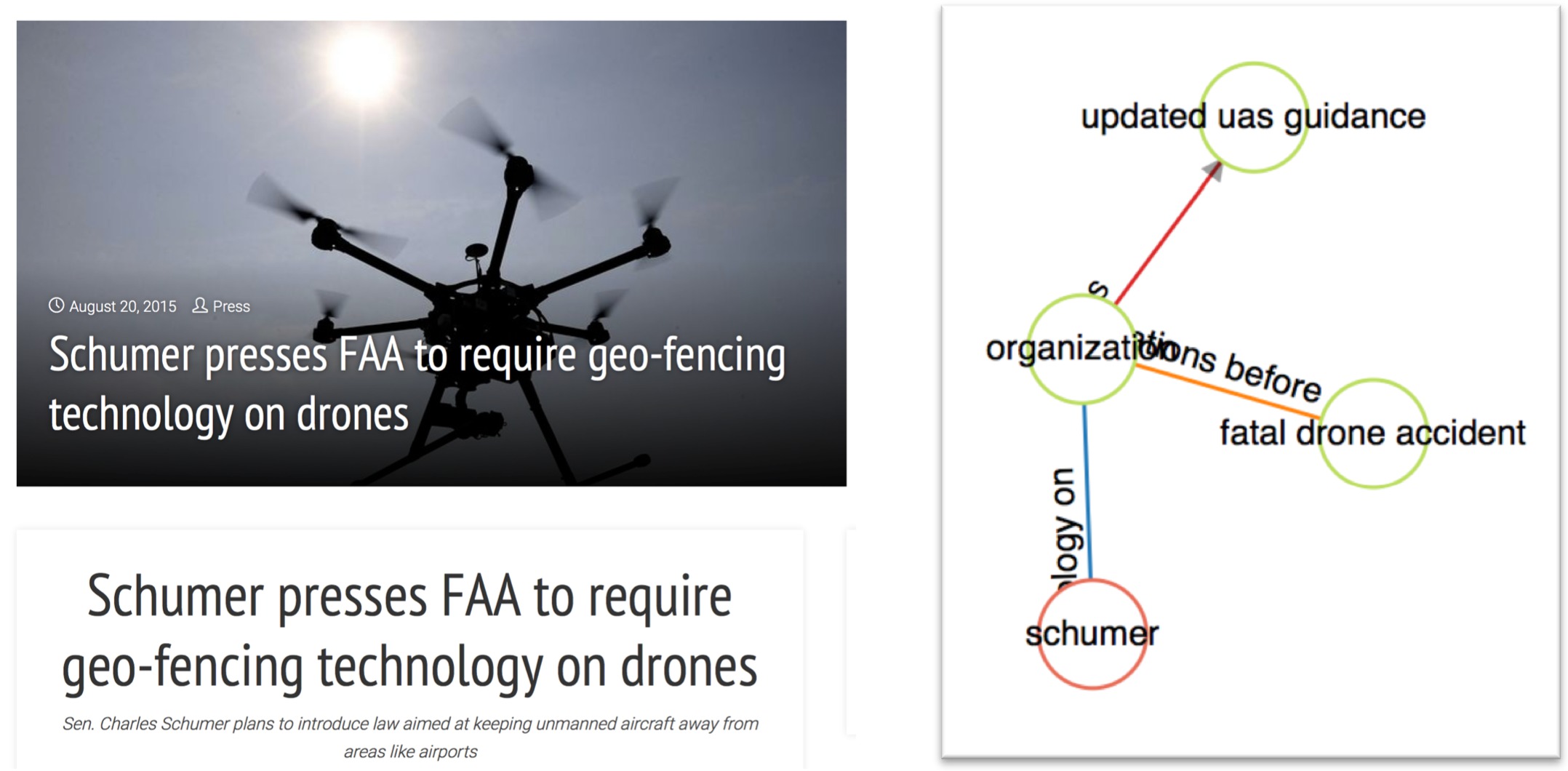}
\caption{Patterns discovered from updates to the knowledge graph.  The updates were learnt from streams of articles obtained from multiple websites.  The example on left demonstrates the validity of the pattern.}
\label{fig:pattern}
\end{figure}

%\begin{figure}[h]
%    \centering
%    \includegraphics[width=0.9\columnwidth]{./figs/HR_yago_K_50}
%    \caption{Illustration of the performance of our initial link prediction algorithm implementation. Benchmarked on a subset of YAGO2 relations, it outperforms state-of-the-art approaches such as matrix factorization for most relations.}
%    \label{fig:linkprediction}
%\end{figure}
%\balancecolumns
\end{document}